\documentclass[manuscript]{aastex}

\shorttitle{Data Exchange Standard for Optical Interferometry}
\shortauthors{Pauls et al.}

\begin{document}

\title{A Data Exchange Standard for Optical\\ (Visible/IR) Interferometry}

\author{T. A. Pauls}
\affil{Naval Research Laboratory, Code 7210, 4555 Overlook Avenue SW, Washington, DC USA 20375-5351}
\email{pauls@nrl.navy.mil\altaffilmark{1}}

\author{J. S. Young}
\affil{Astrophysics Group, Cavendish Laboratory, Madingley Road, CB3 0HE, UK}
\email{jsy1001@cam.ac.uk}

\author{W. D. Cotton}
\affil{NRAO, 520 Edgemont Road, Charlottesville, VA USA 22903}
\email{bcotton@nrao.edu}

\and

\author{J. D. Monnier}
\affil{University of Michigan, Department of Astronomy, Ann Arbor, MI USA 48109}
\email{monnier@umich.edu}

\altaffiltext{1}{We request that comments and suggestions related to the
  OI Exchange Format be directed to the OLBIN email list (see
  \url{http://listes.obs.ujf-grenoble.fr/wws/info/olbin} for
  information on how to subscribe and post to the list).}

\begin{abstract}
  This paper describes the OI Exchange Format, a standard for
  exchanging calibrated data from optical (visible/infrared) stellar
  interferometers. The standard is based on the Flexible Image
  Transport System (FITS), and supports storage of the optical
  interferometric observables including squared visibility and closure
  phase -- data products not included in radio interferometry
  standards such as UV-FITS. The format has already gained the support
  of most currently-operating optical interferometer projects,
  including COAST, NPOI, IOTA, CHARA, VLTI, PTI, and the Keck
  Interferometer, and is endorsed by the IAU Working Group on Optical
  Interferometry. Software is available for reading, writing and
  merging OI Exchange Format files.
\end{abstract}



\section{Introduction}

The idea of a common data format for visible-wavelength and infrared
astronomical interferometry (hence referred to collectively as ``optical
interferometry'') arose through discussions at the June 2000
NSF-sponsored meeting in Socorro~\citep{socorro00}. Since 2001, T.P.
and J.S.Y. have been responsible, under the auspices of the International
Astronomical Union (IAU), for coordinating
discussion on this topic, and for producing and maintaining the
specification document for a format based on the Flexible Image
Transport System (FITS)~\citep{FITS}.

\subsection{Motivation}

The motivation for a creating a data format specific to optical
interferometry was two-fold. Firstly, existing formats designed for
radio interferometry such as UV-FITS~\citep{uvfits} and
FITS-IDI~\citep{fits-idi} do not describe optical data adequately.
Secondly, the more popular UV-FITS format is poorly defined --
its content has evolved and the format is partly defined by reference
to the behaviour of the AIPS software -- and based on the deprecated
FITS ``Random Group'' conventions.

The principal shortcoming of the radio interferometry formats is that
Fourier data is stored as complex visibilities. The noise models for
radio and optical interferometry are quite different; optical
interferometers must usually integrate incoherently to obtain useful
signal-to-noise. Thus the ``good'' observables in the optical are
calibrated fringe power spectra (squared visibility amplitudes) and
bispectra (triple product amplitudes and closure phases), rather than
complex visibilities.  A file format based on complex visibilities can
only encode (optical) closure phases if corresponding visibility
amplitude and closure phase measurements are made simultaneously, and
even then the uncertainties on the closure phases are not stored
(radio interferometry analysis software assumes zero closure phase
error).  There is also no way of saving bispectrum (triple product)
amplitudes.

\subsection{History}

Preliminary drafts of the specification for an optical interferometry
format were discussed by the community at the 198th Meeting of
the American Astronomical Society in June 2001, and the August 2001
Meeting of the IAU Working Group on Optical/IR Interferometry. The
discussion continued by email amongst various interested parties, and
the document was revised.

A public pre-release of the Format Specification and accompanying
example C code was made in March 2002, followed by a second
pre-release of the document in April 2002. This draft was discussed at
the August 2002 IAU Working Group Meeting. Comments were presented by
participants from the European Southern Observatory and the
Interferometry Science Center (now the Michelson Science Center).
Since the IAU Working Group meeting there were two further
pre-releases of the Format Specification, prior to the first
``production'' release.

The standard was frozen on 7 April 2003 (release 5 of the Format
Specification), meaning that subsequent changes would require
increments of the revision numbers for the changed tables. The history
of the Format Specification is summarized in Table~\ref{tab:history}.

The standard is now supported by the majority of optical
interferometer projects, including COAST, NPOI, IOTA, CHARA,
VLTI, PTI, and the Keck Interferometer.

\placetable{tab:history}

\subsection{OIFITS Publications}

This paper contains the formal definition of the standard, in
Sec.~\ref{sec:start} onwards.  As earlier versions of the Format
Specification have only been distributed via the world-wide-web, this
paper formalises the standard and serves as a published reference for
it. Further discussion explaining the design decisions and providing
explicit pointers to existing software for reading and writing the
format can be found in \citet{oifits04}, which is intended as a
``companion'' to the official Format Specification given here.

A first application of the Exchange Format, a ``beauty contest'' to
compare the performance of a number of image reconstruction software
packages, is described by \citet{beauty04}. The contest demonstrated
that four different international groups could faithfully reconstruct
images from simulated datasets in the OIFITS format.

No alterations have been made to the standard since release 5 of the
Format Specification. The text in Sec.~\ref{sec:start} onwards is
substantially the same as that in release 5, apart from minor changes
to clarify certain points and to conform to the journal style. The
revision numbers of all tables defined by the standard remain at
unity.

%

\section{Purpose and Scope} 
\label{sect:purpose}

By defining and maintaining a common data format, we hope to encourage the
development of common software for analysis of data from optical
stellar interferometers, and to facilitate the exchange of data
between different research groups. In this way, the limited Fourier
coverage of current instruments can be ameliorated by combining data
from several interferometers. An example of this is given in the
paper by \citet{keckiota}.

The format is intended to support the storage of calibrated,
time-averaged data. Such data can be prepared from the raw fringe
measurements without using information about the detailed structure of
the target (i.e.\ without doing any astrophysical interpretation), yet
once the data is in the format, it can be analysed without knowing the
details of the instrument. Calibrated data from different
interferometers can be treated in the same way (provided there are no
residual systematic errors).

The standard includes the information needed to perform ``imaging''
from a calibrated data-set. Here imaging refers loosely to any process
for making inferences about the sky brightness distribution from the
data. The standard explicitly allows extra binary tables or columns,
beyond those it defines, to be used to store additional information.
In this way, the standard can be used for other purposes besides
imaging, e.g.\ for astrometry or as an archive format.

We suggest that the common data format be referred to as the ``OI
Exchange Format'', or the ``Exchange Format'' when the context is
clear. A suitable very short moniker is ``OIFITS'' (or ``OI-FITS'').


\section{Document Conventions}
\label{sec:start}

In what follows we use the FITS binary table nomenclature of keywords
and column headings. The values associated with the keywords can be
considered as scalars, while each column can be simply an array, or an
array of ``pointers'' to other arrays. The following data types are
used in the standard: I = integer (16-bit), A = character, E = real
(32-bit), D = double (64-bit), L = logical. In the tables below, the
number in parentheses is the dimensionality of the entry.  The table
names given below correspond to the values of the EXTNAME keyword.
Other mandatory keywords describing the structure of the FITS binary
tables~\citep[see][]{FITS} have been omitted from this document.
\citet{FITS} also describes various extensions to binary tables that
are not part of the FITS standard.  None of these are currently used
in this format.  The definitions of all tables have been ``frozen''
since April 2003. All revision numbers are currently equal to one. Any
future changes will require increments in the revision numbers of the
altered tables.


\section{Definitions and Assumptions}
\label{sec:defn}

The conventions described here are generally identical to those used
in radio interferometry and described in standard
textbooks~\citep[e.g.\ ][]{TMS}.

\subsection{Baseline vector}

The baseline vector between two interferometric stations A and B whose
position vectors are $\vec{x_A}$ and $\vec{x_B}$ is
defined as $\vec{AB} = \vec{x_B} - \vec{x_A}$.

\subsection{UV coordinates}

$u$ is the East component and $v$ is the North component of the
projection of the baseline vector onto the plane normal to the
direction of the phase center, which is assumed to be the pointing
center.

\subsection{Complex Visibility}

The basic observable of an interferometer is the complex visibility,
which is related to the sky brightness distribution $I(x,y)$ by a
Fourier Transform:
\begin{equation} \label{eqn:vis}
V(u, v) = \int \!\! \int \mathrm{d}x\, \mathrm{d}y\,\, I(x,y)\exp{(-2\pi i (ux+vy))} \, .
\end{equation}
$x$ and $y$ are displacements (in radians) in the plane of the sky
(which is assumed to be flat). The origin of this coordinate system is
the phase center. $x$ is in the direction of increasing right
ascension (i.e. the $x$ axis points East), and $y$ is in the direction
of increasing declination (the $y$ axis points North).
With $x$ and $y$ defined, the above equation defines the sign
convention for complex visibilities that should be used in the data
format.  The visibility is normalized by the total flux from the
target, which is assumed to remain constant over the time spanned by
the measurements in the file. Neither the field of view over which the
``total'' flux is collected, or the field of view over which fringes
are detected (i.e. the limits of the above integral), can be inferred
from the data file.

\subsection{Squared Visibility}

The squared visibility is simply the modulus squared of the complex visibility:
\begin{equation} \label{eqn:vis2}
S(u, v) = {\left| V(u, v) \right|}^2 \, .
\end{equation}

\subsection{Triple Product}

The triple product, strictly the bispectrum, is the product of the
complex visibilities on the baselines forming a closed loop joining
three stations A, B and C. In the following expression, $(u_1, v_1)$
is the projection of AB and $(u_2, v_2)$ is the projection of BC (and
hence $(u_1+u_2, v_1+v_2)$ is the projection of AC):
\begin{equation} \label{eqn:t3}
T(u_1, v_1, u_2, v_2) = V(u_1, v_1)\, V(u_2, v_2)\, V^*(u_1+u_2,v_1+v_2) \, .
\end{equation}
 
\subsection{Noise model for triple product}

\begin{figure}
  \centering
  \includegraphics[width=10cm]{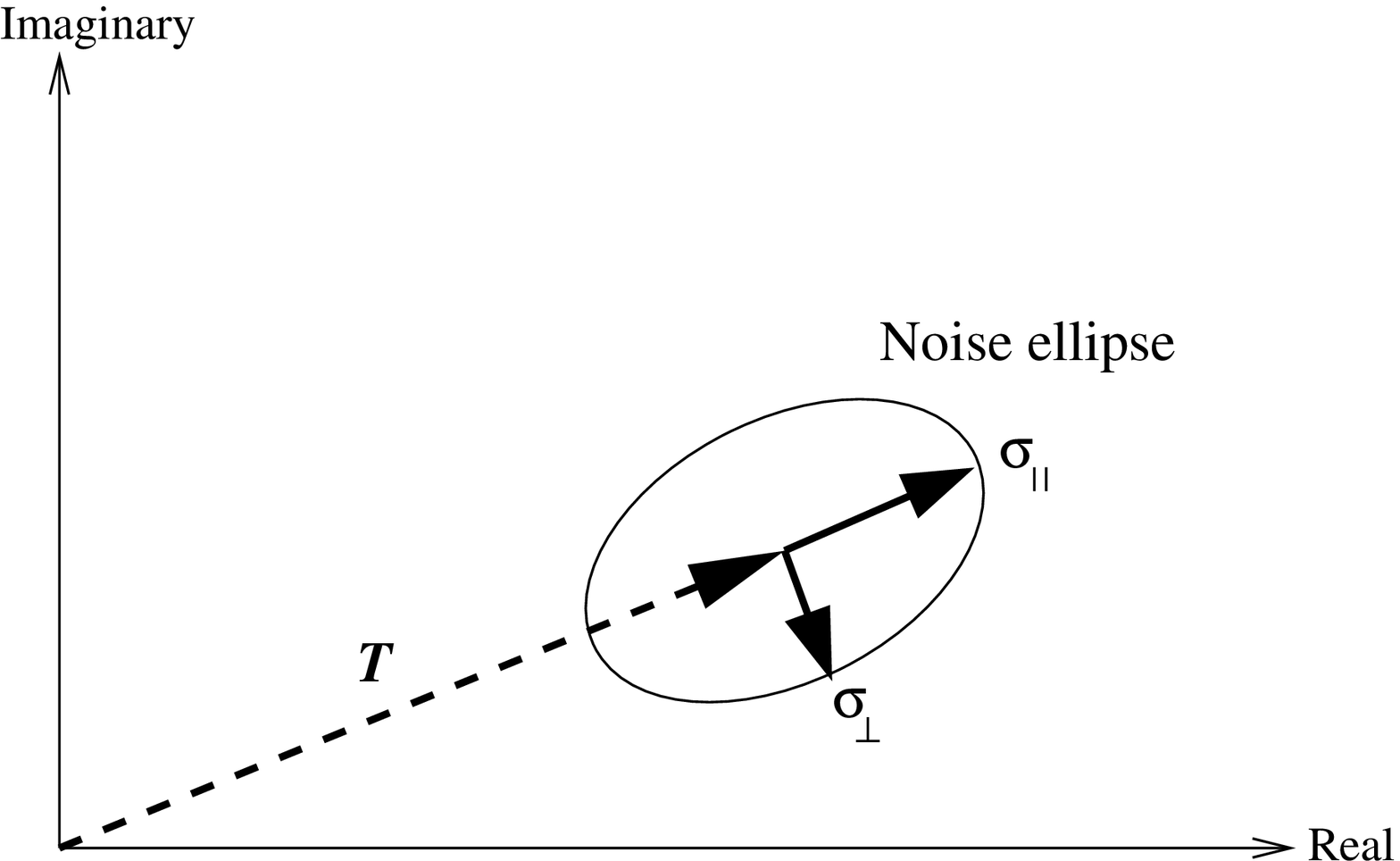}
  \caption{\label{fig:t3noise} Noise model for triple product. The
    figure is an Argand diagram showing the mean triple product $T$
    with its associated noise ellipse, elongated parallel to $T$. See
    text for explanation.}
\end{figure}

The data are assumed to be complex triple products averaged over a
large number of ``exposures''. In such a case, the noise can be fully
described in terms of a Gaussian noise ellipse in the complex plane.
Photon, detector and background noise tend to lead to noise ellipses
that are close to circular. On the other hand, fluctuating atmospheric
phase errors across telescope apertures typically cause fluctuations
in the amplitude of the triple product which are much larger than the
fluctuations in the phase. Thus the ``atmospheric'' contribution to
the noise ellipse is elongated along the direction of the mean triple
product vector in the Argand diagram, as shown in
Fig.~\ref{fig:t3noise}. Such noise needs to be characterised in terms
of the variance $\sigma_\perp^2$ perpendicular to the mean triple
product vector and the variance $\sigma_\parallel^2$ parallel to $T$.
We can parameterize the perpendicular variance in terms of a ``phase
error'' $\sigma_\theta = (180/\pi)(\sigma_\perp/|T|)$. The phase
error gives an approximate value for the rms error in the closure
phase in degrees. We denote $\sigma_\parallel$ as the ``amplitude
error''.

\placefigure{fig:t3noise}

In many cases, the observer may be interested primarily in the closure
phase and not the triple product amplitude, and therefore may choose
not to calibrate the amplitude. Such a case can be indicated in the
above notation as an infinite amplitude error and a finite phase
error. The data format specifies that such a case should be indicated
by a NULL value for the amplitude (the amplitude error value is then
ignored).

\subsection{Noise model for complex visibility}

There was much discussion on the (now defunct)
\texttt{oi-data@rsd.nrl.navy.mil} email list of the representation to
use for complex visibilities in the standard. A number of different
classes of data can be represented as complex visibilities, including
several varieties of differential phase data.  In all cases the
standard should only be used to store averaged data.  Thus, as with
triple products, we must consider the shape of the noise ellipse in
the complex plane.

It has been demonstrated~\citep{complexvis} that
both circularly-symmetric noise, and noise ellipses elongated parallel
to or perpendicular to the mean vector can occur in practice. Thus far
there has been no evidence for noise ellipses elongated parallel to
the real or imaginary axes, although examples of some classes of data
have yet to be presented.  Hence an amplitude/phase representation of
complex visibilities, mirroring that used for triple products, has
been adopted in the current version of the standard.


\section{FITS File Structure}

A valid exchange-format FITS file must contain one (and only one)
OI\_TARGET table, plus one or more of the data tables: OI\_VIS,
OI\_VIS2, or OI\_T3. Each data table must refer to an OI\_WAVELENGTH
table that is present in the file. There may be more than one of each
type of data table (e.g. OI\_VIS2). One or more OI\_ARRAY tables (or
equivalent e.g.  for aperture masking, in future releases of the
standard) may optionally be present. Where multiple tables of the same
EXTNAME are present, each should have a unique value of EXTVER (this
according to the FITS standard -- however the example C code and
J.D.M.'s IDL software do not require EXTVER to be present).

The tables can appear in any order. Other header-data units may appear
in the file, provided their EXTNAMEs do not begin with ``OI\_''.
Reading software should not assume that either the keywords or the
columns in a table appear in a particular order. This is
straightforward to implement using software libraries such as cfitsio.

Any of the tables may have extra keywords or columns beyond those
defined in the standard. It would facilitate the addition of new
keywords and columns in future releases of the standard if the
non-standard keywords and column names were given a particular prefix
e.g. ``NS\_'', to avoid conflicts.

\section{Tables Defined by the Standard}

\subsection{OI\_ARRAY (revision 1)}

As defined, this table is aimed at ground-based interferometry with
separated telescopes.  Alternative tables could be used for other
cases. These must have at least an ARRNAME keyword, for
cross-referencing purposes. Each OI\_ARRAY-equivalent table in a file
must have a unique value for ARRNAME.

\subsubsection*{Keywords}

\begin{tabular}{lll}
OI\_REVN & I & Revision number of the table definition\\
ARRNAME  & A & Array name, for cross-referencing\\
FRAME    & A & Coordinate frame\\
ARRAYX   & & \\
ARRAYY   & D & Array center coordinates (meters)\\
ARRAYZ   & &
\end{tabular}

\subsubsection*{Column Headings (one row for each telescope)}

\begin{tabular}{lll}
TEL\_NAME  & A (16) & Telescope name\\
STA\_NAME  & A (16) & Station name\\
STA\_INDEX & I (1)  & Station number\\
DIAMETER   & E (1)  & Element diameter (meters)\\
STAXYZ     & D (3)  & Station coordinates relative to array center (meters)
\end{tabular}

\subsubsection*{Number of elements}

There is no keyword giving the number of elements (NELEMENT in a
previous revision of this document), as this is equal to the number of
rows in the FITS binary table, which is given by the standard NAXIS2
keyword. For the same reason, there are no format- specific keywords
giving the number of rows in any of the other tables.

\subsubsection*{Coordinate frame}

If the FRAME keyword has the value ``GEOCENTRIC'', then the
coordinates are given in an earth-centered, earth-fixed, Cartesian
reference frame. The origin of the coordinates is the earth's centre
of mass. The $z$ axis is parallel to the direction of the conventional
origin for polar motion. The $x$ axis is parallel to the direction of
the intersection of the Greenwich meridian with the mean astronomical
equator. The $y$ axis completes the right- handed, orthogonal
coordinate system.

Currently, no other values for the FRAME keyword may be used. This
will change if the need arises.

\subsubsection*{Array coordinates}

The ARRAYX, ARRAYY, and ARRAYZ keywords shall give the coordinates of
the array center in the coordinate frame specified by the FRAME
keyword. Element coordinates in the main part of the table are given
relative to the array center, in the same coordinate frame.
Coordinates are given in meters.

\subsubsection*{Station number}

Each row in the table shall be assigned a unique station number, which
shall be used in other tables as an index into this one.  The table
structure is the simplest possible i.e. there is no explicit concept
of different ``configurations'' within the table. Each row in the
table shall correspond to a distinct set of station coordinates used
in taking the data stored in the file.

\subsubsection*{Element diameter}

This is the effective aperture size, e.g. if the telescope is stopped down.

\subsection{OI\_TARGET (revision 1)}

\subsubsection*{Keywords}
\begin{tabular}{lll}
        OI\_REVN & I & Revision number of the table definition
\end{tabular}

\subsubsection*{Column Headings (one row for each source)}
\begin{tabular}{lll}
        TARGET\_ID & I (1) & Index number\\
        TARGET   & A (16) & Target name\\
        RAEP0    & D (1) & RA at mean equinox (degrees)\\
        DECEP0   & D (1) & DEC at mean equinox (degrees)\\
        EQUINOX  & E (1) & Equinox\\
        RA\_ERR  & D (1) & Error in RA at mean equinox (degrees)\\
        DEC\_ERR & D (1) & Error in DEC at mean equinox (degrees)\\
        SYSVEL   & D (1) & Systemic radial velocity (meters per second)\\
        VELTYP   & A (8) & Reference for radial velocity (``LSR'', ``GEOCENTR'', etc.)\\
        VELDEF   & A (8) & Definition of radial velocity (``OPTICAL'', ``RADIO'')\\
        PMRA     & D (1) & Proper motion in RA (degrees per year)\\
        PMDEC    & D (1) & Proper motion in DEC (degrees per year)\\
        PMRA\_ERR  & D (1)  & Error of proper motion in RA (degrees per year)\\
        PMDEC\_ERR & D (1)  & Error of proper motion in DEC
 (degrees per year)\\
        PARALLAX   & E (1)  & Parallax (degrees)\\
        PARA\_ERR & E (1)  & Error in parallax (degrees)\\
        SPECTYP    & A (16) & Spectral type
\end{tabular}

\subsubsection*{Target position}

The RAEP0 and DECEP0 columns shall contain the right ascension and
declination respectively of the phase center at the standard mean
epoch, in degrees. RA\_ERR and DEC\_ERR shall contain the one-sigma
uncertainties in these quantities.  The phase center is assumed to be
the pointing center. The EQUINOX field shall contain a (floating
point) Julian year giving both the epoch of the position (RAEP0,
DECEP0) and the equinox for the celestial coordinate system in which
the position is expressed.  The PMRA and PMDEC columns should contain
the proper motions of the source in right ascension and declination
respectively, in degrees per Julian year. If the proper motion is
unknown then both fields should be set to zero. PMRA\_ERR and
PMDEC\_ERR shall contain the one-sigma uncertainties in these
quantities.  If an apparent position at the time of an observation is
required, it should be obtained by applying the appropriate
transformations to the catalogue position given by RAEP0 and DECEP0,
making use of PMRA, PMDEC, and PARALLAX.

\subsubsection*{Velocity Information}

The SYSVEL column shall give the systemic radial velocity of the
target (positive if receding).  The VELTYP column shall contain a
string that specifies the frame of reference for the systemic
velocities. The string shall be one of the following:

\begin{tabular}{ll}
LSR & Local Standard of Rest\\
HELIOCEN & Relative to the SUN\\
BARYCENT & Solar system barycenter\\
GEOCENTR & Center of mass of the earth\\
TOPOCENT & Uncorrected\\
\end{tabular}

The VELDEF column shall contain a string indicating the convention
used for the (relativistic) systemic velocities. It shall be either
``RADIO'' or ``OPTICAL'' (the distinction is not important for
velocities much less than the speed of light).

\subsection{OI\_WAVELENGTH (revision 1)}

\subsubsection*{Keywords}

\begin{tabular}{lll}
        OI\_REVN & I & Revision number of the table definition\\
        INSNAME  & A & Name of detector, for cross-referencing
\end{tabular}

\subsubsection*{Column Headings (one row for each wavelength channel)}

\begin{tabular}{lll}
        EFF\_WAVE & E & Effective wavelength of channel (meters)\\
        EFF\_BAND & E & Effective bandpass of channel (meters)
\end{tabular}

\subsubsection*{Name of detector}

Each OI\_WAVELENGTH table in a file must have a unique value for
INSNAME.

\subsubsection*{Wavelengths}

Each OI\_WAVELENGTH table describes the spectral
response of detector(s) with a number of spectral channels. Each table
gives the wavelengths for one or more of the data tables (OI\_VIS,
OI\_VIS2, OI\_T3), and will often correspond to a single physical
detector.

The EFF\_WAVE column shall contain the best available estimate of the
effective wavelength of each spectral channel, and the EFF\_BAND
column shall contain the best available estimate of the effective
half-power bandwidth. These estimates should include the effect of the
earth's atmosphere, but not the spectrum of the target (the effect of
the target spectrum should be taken into account as part of any
model-fitting/mapping process, i.e. the target spectrum is part of the
model).

\subsection{OI\_VIS (revision 1)}

\subsubsection*{Keywords}

\begin{tabular}{lll}
      OI\_REVN  & I & Revision number of the table definition\\
      DATE-OBS  & A & UTC start date of observations\\
      ARRNAME   & A & (optional) Identifies corresponding OI\_ARRAY\\
      INSNAME   & A & Identifies corresponding OI\_WAVELENGTH table\\
\end{tabular}

\subsubsection*{Column Headings (one row for each measurement)}

\begin{tabular}{lll}
      TARGET\_ID & I (1) & Target number as index into OI\_TARGET table\\
      TIME       & D (1) & UTC time of observation (seconds)\\
      MJD        & D (1) & Modified Julian Day\\
      INT\_TIME  & D (1) & Integration time (seconds)\\
      VISAMP     & D (NWAVE) & Visibility amplitude\\
      VISAMPERR  & D (NWAVE) & Error in visibility amplitude\\
      VISPHI     & D (NWAVE) & Visibility phase in degrees\\
      VISPHIERR  & D (NWAVE) & Error in visibility phase in degrees\\
      UCOORD     & D (1) & U coordinate of the data (meters)\\
      VCOORD     & D (1) & V coordinate of the data (meters)\\
      STA\_INDEX & I (2) & Station numbers contributing to the data\\
      FLAG       & L (NWAVE) & Flag
\end{tabular}

\subsection{OI\_VIS2 (revision 1)}

\subsubsection*{Keywords}

\begin{tabular}{lll}
      OI\_REVN  & I & Revision number of the table definition\\
      DATE-OBS  & A & UTC start date of observations\\
      ARRNAME   & A & (optional) Identifies corresponding OI\_ARRAY\\
      INSNAME   & A & Identifies corresponding OI\_WAVELENGTH table\\
\end{tabular}

\subsubsection*{Column Headings (one row for each measurement)}

\begin{tabular}{lll}
      TARGET\_ID & I (1)     & Target number as index into OI\_TARGET table\\
      TIME       & D (1)     & UTC time of observation (seconds)\\
      MJD        & D (1)     & Modified Julian Day\\
      INT\_TIME  & D (1)     & Integration time (seconds)\\
      VIS2DATA   & D (NWAVE) & Squared Visibility\\
      VIS2ERR    & D (NWAVE) & Error in Squared Visibility\\
      UCOORD     & D (1)     & U coordinate of the data (meters)\\
      VCOORD     & D (1)     & V coordinate of the data (meters)\\
      STA\_INDEX & I (2)     & Station numbers contributing to the data\\
      FLAG       & L (NWAVE) & Flag
\end{tabular}

\subsection{OI\_T3 (revision 1)}

\subsubsection*{Keywords}

\begin{tabular}{lll}
      OI\_REVN  & I & Revision number of the table definition\\
      DATE-OBS  & A & UTC start date of observations\\
      ARRNAME   & A & (optional) Identifies corresponding OI\_ARRAY\\
      INSNAME   & A & Identifies corresponding OI\_WAVELENGTH table
\end{tabular}

\subsubsection*{Column Headings (one row for each measurement)}
\begin{tabular}{lll}
      TARGET\_ID & I (1) & Target number as index into OI\_TARGET table\\
      TIME       & D (1) & UTC time of observation (seconds)\\
      MJD        & D (1) & Modified Julian Day\\
      INT\_TIME  & D (1) & Integration time (seconds)\\
      T3AMP      & D (NWAVE) & Triple Product Amplitude\\
      T3AMPERR   & D (NWAVE) & Error in Triple Product Amplitude\\
      T3PHI      & D (NWAVE) & Triple Product Phase in degrees\\
      T3PHIERR   & D (NWAVE) & Error in Triple Product Phase in degrees\\
      U1COORD    & D (1) & U coordinate of baseline AB of the triangle (meters)\\
      V1COORD    & D (1) & V coordinate of baseline AB of the triangle (meters)\\
      U2COORD    & D (1) & U coordinate of baseline BC of the triangle (meters)\\
      V2COORD    & D (1) & V coordinate of baseline BC of the triangle (meters)\\
      STA\_INDEX & I (3) & Station numbers contributing to the data\\
      FLAG       & L (NWAVE) & Flag
\end{tabular}

\vspace{2ex}
\noindent The following comments apply to one or more of the OI\_VIS, OI\_VIS2,
and OI\_T3 tables.

\subsubsection*{Cross-referencing}

Each data table must refer (via the INSNAME keyword) to a particular
OI\_WAVELENGTH table describing the wavelength channels for the
measurements.  Each data table may optionally refer, via the ARRNAME
keyword, to an OI\_ARRAY table.

\subsubsection*{Start date of observations}
This shall be a UTC date in the format YYYY-MM-DD, e.g.\ ``1997-07-28''.

\subsubsection*{Time of observation}

The value in the TIME column shall be the mean UTC time of the
measurement in seconds since 0h on DATE-OBS. Note this may take
negative values, or values $> 86400$ seconds, and hence the epoch of
observation for the particular table is not restricted to DATE-OBS.

The value in the MJD column shall be the mean UTC time of the
measurement expressed as a modified Julian Day. It might be
appropriate to use the MJD values instead of the TIME values when
dealing with long time-spans, but the standard makes no stipulation in
this regard.

\subsubsection*{Integration time}

The exchange format will normally be used for interchange of
time-averaged data. The ``integration time'' is therefore the length
of time over which the data were averaged to yield the given data
point.

\subsubsection*{Data arrays}

If the triple product amplitudes are meaningless, as is the case for
COAST data, NULL values for T3AMP may be used. The closure phases
should still be treated as valid.

NWAVE is the number of distinct spectral channels recorded by the
single (possibly ``virtual'') detector, as given by the NAXIS2 keyword
of the relevant OI\_WAVELENGTH table.

\subsubsection*{Complex visibility and visibility-squared UV
  coordinates}

UCOORD, VCOORD give the coordinates in meters of the point in the UV
plane associated with the vector of visibilities. The data points may
be averages over some region of the UV plane, but the current version
of the standard says nothing about the averaging process. This may
change in future versions of the standard.

\subsubsection*{Triple product UV coordinates}

The U1COORD, V1COORD, U2COORD, and V2COORD columns contain the
coordinates of the bispectrum point -- see Sec.~\ref{sec:defn} for
details. Note that U3COORD and V3COORD are implicit.

The corresponding data points may be averages in (bi-) spatial
frequency space, but this version of the standard does not attempt to
describe the averaging process.

\subsubsection*{Flag}

If a value in this vector is true, the corresponding datum should be
ignored in all analysis.

\subsection{Optional Tables}

It may be useful to allow for some optional tables. For example, there
might be one that contains instrument specific information, such as
the backend configuration. Another optional table could contain
information relevant to astrometry.  The EXTNAMEs of additional tables
should not begin with ``OI\_''.


\acknowledgments
Development of the format was performed under the auspices of the
IAU Working Group on Optical/IR Interferometry, and has the strong support of
its members. We would like to thank Peter~R. Lawson (Chair of the
Working Group) for his encouragement and support.

The authors would like to thank David Buscher, Christian Hummel, and David
Mozurkewich for providing material for the Format Specification and the
OIFITS website. We would like to take this opportunity to thank all
those in the community who have contributed to the definition of the
format and to its take-up by many interferometer projects.



\clearpage


\clearpage

\begin{deluxetable}{ll}
  \tabletypesize{\tiny}
  \tablecaption{\label{tab:history} History of the OI Exchange Format.}
  \tablehead{
    \colhead{Date} & \colhead{Description}}
  \startdata
  2002/03/25 & Pre-release of document and example code -- Release 1 of Format Specification\\
  2002/04/25 & Minor correction to document -- Release 2 of Format Specification\\
  2002/11/26 & Post-IAU-WG-meeting release of document and example -- Release 3 of Format Specification\\
  2003/02/17 & Release of document, C and IDL code to fix problem in OI\_TARGET table -- Release 4 of Format Specification\\
  2003/04/07 & Freeze of format specification. Revision numbers of all tables set to unity -- Release 5 of Format Specification\\
  2005       & This paper, no changes to Exchange Format or table revision numbers\\
  \enddata
\end{deluxetable}


\begin{thebibliography}{9}
\expandafter\ifx\csname natexlab\endcsname\relax\def\natexlab#1{#1}\fi

\bibitem[{Cotton {et~al.}(1990)}]{uvfits}
Cotton, W.~D., {et~al.} 1990, Going AIPS: A Programmer's Guide to the NRAO
  Astronomical Image Processing System (National Radio Astronomy Observatory,
  Charlottesville, VA), 14.7--14.10,
  \url{http://www.aoc.nrao.edu/aips/goaips.html}

\bibitem[{Flatters(2000)}]{fits-idi}
Flatters, C. 2000, AIPS Memo No. 102: The {FITS} Interferometry Data
  Interchange Format (National Radio Astronomy Observatory)

\bibitem[{Hanisch {et~al.}(2001)Hanisch, Farris, Greisen, Pence, Schlesinger,
  Teuben, Thompson, \& Warnock}]{FITS}
Hanisch, R.~J., Farris, A., Greisen, E.~W., Pence, W.~D., Schlesinger, B.~M.,
  Teuben, P.~J., Thompson, R.~W., \& Warnock, A. 2001, \aap, 376, 359

\bibitem[{Hummel {et~al.}(2002)Hummel, Pauls, \& Mozurkewich}]{complexvis}
Hummel, C.~A., Pauls, T.~A., \& Mozurkewich, D. 2002,
  \url{http://www.mrao.cam.ac.uk/\%7Ejsy1001/exchange/complex/complex.html}

\bibitem[{Lawson {et~al.}(2004)Lawson, Cotton, Hummel, Monnier, Zhao, Young,
  Thorsteinsson, Meimon, Mugnier, Besnerais, Thi{\'e}baut, \&
  Tuthill}]{beauty04}
Lawson, P.~R., Cotton, W.~D., Hummel, C.~A., Monnier, J.~D., Zhao, M., Young,
  J.~S., Thorsteinsson, H., Meimon, S.~C., Mugnier, L., Besnerais, G.~L.,
  Thi{\'e}baut, E., \& Tuthill, P.~G. 2004, in \procspie, Vol. 5491, New
  Frontiers in Stellar Interferometry, ed. W.~Traub, J.~D. Monnier, \&
  M.~Sch{\"o}ller, 21--25~Jun.~2004, Glasgow (SPIE Press), 886

\bibitem[{McAlister \& Cornwell(2000)}]{socorro00}
McAlister, H., \& Cornwell, T., eds. 2000, Report on the Workshop on Imaging
  with Ground-based optical interferometers, National Science Foundation,
  \url{http://olbin.jpl.nasa.gov/papers/Report1.0.PDF}

\bibitem[{Monnier {et~al.}(2004)Monnier, Millan-Gabet, Tuthill, Traub,
  Carleton, du~Foresto, Danchi, Lacasse, Morel, Perrin, Porro, Schloerb, \&
  Townes.}]{keckiota}
Monnier, J., Millan-Gabet, R., Tuthill, P., Traub, W., Carleton, N.,
  du~Foresto, V.~C., Danchi, W., Lacasse, M., Morel, S., Perrin, G., Porro, I.,
  Schloerb, F., \& Townes., C. 2004, \apj, 605, 436

\bibitem[{Pauls {et~al.}(2004)Pauls, Young, Cotton, \& Monnier}]{oifits04}
Pauls, T.~A., Young, J.~S., Cotton, W.~D., \& Monnier, J.~D. 2004, in
  \procspie, Vol. 5491, New Frontiers in Stellar Interferometry, ed. W.~Traub,
  J.~D. Monnier, \& M.~Sch{\"o}ller, 21--25~Jun.~2004, Glasgow (SPIE Press),
  1231

\bibitem[{Thompson {et~al.}(1986)Thompson, Moran, \& Swenson~Jr.}]{TMS}
Thompson, A.~R., Moran, J.~M., \& Swenson~Jr., G.~W. 1986, Interferometry and
  Aperture Synthesis in Radio Astronomy (Wiley Interscience)

\end{thebibliography}
\end{document}